\documentclass[twocolumn,showpacs,prl,floatfix]{revtex4}

\usepackage{graphicx}
\usepackage{bm}
\begin{document}

\title{Phonon Hall thermal conductivity from Green-Kubo formula}

\author{Jian-Sheng Wang}
 \altaffiliation[Also at Institute of High Performance Computing, 
1 Fusionopolis Way, {\#}16-16 Connexis, Singapore 138632, and Singapore-MIT Alliance, 4 Engineering Drive 3, Singapore 117576.]
\and
\author{Lifa Zhang}%
\affiliation{Department of Physics and Center for Computational Science and Engineering, National University of Singapore, Singapore 117542, Republic of Singapore}

\date{7 February 2009}

\begin{abstract}
We derive a formula for the thermal conductivity tensor of a ballistic phonon
Hall model.  It is found that, although the diagonal elements of the
conductivity tensor diverge to infinite, the off-diagonal elements are finite,
antisymmetric, and odd in magnetic field.  The off-diagonal elements are
non-zero only if the dynamic matrix of the phonon system breaks mirror
reflection symmetry.  The results are obtained without perturbative
assumptions about the spin-phonon interactions.
\end{abstract}

\pacs{66.70.-f, 72.10.Bg, 72.15.Gd, 72.20.Pa}
\maketitle


The Hall effect of electronic conduction is well-known and has many
applications.  The analogous effects for the transport of gas molecules, 
spins, and photons also exit \cite{beenakker67,sczhang03,photo08}.  The
phonon Hall effect, that is, the appearance of a transverse thermal current
when a magnetic field is applied perpendicular to the direction of temperature
gradient, is esoteric and not well understood. Electrons couple directly to
the magnetic field through the Lorentz force.  There is no obvious coupling
between phonons and magnetic field.  In 2005, Strohm {\sl et al.}\ reported
such an effect in a paramagnetic dielectric garnet Tb$_3$Ga$_5$O$_{12}$
\cite{strohm05} and confirmed also in Ref.~\cite{inyushkin07}, and called it
phonon Hall effect.  Two theoretical papers followed
\cite{sheng06,kagan08}. Both of them considered a similar model of
the spin-phonon interaction, and both of them treated the interaction
perturbatively. The work of Ref.~\cite{kagan08} appears to imply that
ballistic systems cannot produce a phonon Hall effect, and the authors evoked further
high order spin-phonon interaction terms to demonstrate the existence of the
effect.  Although the two approaches are quite different, one based on
Green-Kubo formula, the other on Boltzmann-type kinetic equation, curiously,
the final results for the off-diagonal thermal conductivity tensor are
similar.

In this paper, we address the following issues. (1) Is a ballistic
system capable of producing the phonon Hall effect?  Our answer to this
question is affirmative.  Although the effect will be smaller as the linear
size $L$ of the system becomes larger (scaled as $1/L$).  (2) What is the role
of symmetry?  We found that break of a mirror reflection symmetry is essential
to observe the phonon Hall effect.  If the system looks the same inside a
mirror, we should not observe such effect on very general ground.  We use the
same model as that of Refs.~\cite{sheng06,kagan08} but without the
perturbative assumption.  The perturbative expansion with respect to the
spin-phonon interaction breaks down near the $\Gamma$-point of the phonon
dispersion.  This complicates the behavior of thermal conductivity at very low
temperatures.  Since the model is ballistic, the thermal conductivity in
general should diverge with the system sizes. But for isotropic systems like
the two-dimensional square or honeycomb lattices, the off-diagonal thermal
conductivity is in fact finite.  In the following, we introduce the model, outline
a derivation of the thermal conductivity using Green-Kubo formula, and present
numerical results and give some comparison with experiments.

We consider a harmonic periodic lattice with the extra Rahman (or spin-orbit)
interaction at each lattice site proportional to ${\bf s} \cdot ({\bf r}
\times {\bf p})$.  Here ${\bf s}$ is the (pseudo-) spin representing the
Kramer doublet; $\bf r$ and $\bf p$ are displacement and conjugate momentum.
We'll replace $\bf s$ by an average magnetization of the system and choose the
vector to be in $z$ direction.  The explicit spin degrees of freedom drop out
of the problem.  The Hamiltonian of the system can be written in a compact
form
\begin{equation}
H = \frac{1}{2} p^T p + \frac{1}{2} u^T K u + u^T\! A\, p,
\label{eq-model}
\end{equation}
where $u$ is a column vector of displacements away from lattice equilibrium
positions for all the degrees of freedoms, multiplied by $\sqrt{m}$, $p$ is
the associated conjugate momenta.  The Rahman term, $u^TA p$, is onsite; 
the matrix $A$ is an antisymmetric real matrix, $A^T = - A$, and is  
block diagonal with
diagonal elements (in two dimensions)
\begin{equation}
\left(\begin{array}{rr} 0 & +h  \\
         -h & 0  \\
       \end{array}\right).
\end{equation}
We'll call $h$ magnetic field although $h$ is only proportional to the real
magnetic field in a paramagnet. It has the dimension of frequency.  Since the
interaction term depends on momentum, the velocity and momentum are not the
same but related through $\dot{u} = p - Au$.  This is the same model studied
in Refs.~\cite{sheng06,kagan08} except a slightly different notation.  It has
been proposed (in a more general form) based on quantum theory and fundamental
symmetries long time ago to study spin-phonon interactions
\cite{old-sp-papers,sp-book,ioselevich95}.

Equation~(\ref{eq-model}) is quadratic in the dynamic variables $u$ and $p$,
thus is amenable for an exact solution.  Our calculation procedure is as
follows.  We first obtain the eigen modes of the system.  Using the eigen
modes, we give an expression for the energy current.  We then apply the
Green-Kubo formula to compute the thermal conductivity tensor.  Since the
system is periodic, we can apply the Bloch theorem.  The polarization vector
$\epsilon$ then satisfies
\begin{equation}
\bigl[ (-i\omega + A)^2 + D\bigr] \epsilon = 0,
\label{eq-disper1}
\end{equation}
where $D({\bf k}) = \sum_{l'} K_{l,l'} e^{i({\bf R}_{l'} - {\bf R}_{l})\cdot
{\bf k}}$ is the dynamic matrix.  $K_{l,l'}$ is the submatrix between unit
cell $l$ and $l'$ in the full spring constant matrix $K$; ${\bf R}_l$ is the
real-space lattice vector.  This equation is not a standard eigenvalue
problem.  It is numerically more advantageous to consider both the coordinates
and momenta and to solve an eigenvalue problem:
\begin{equation}
i\omega\, x = \left( \begin{array}{cc} A & D \\
                      -I & A \end{array} \right) x,
\label{eq-eigen}
\end{equation}
where $x = (\mu, \epsilon)^T$ and $I$ is an identity matrix. Contrary to the
usual lattice dynamic problems, the polarization vectors are not orthogonal to
each other.  We need to consider both the right and left eigen vectors.
Because of the special form of Eq.~(\ref{eq-eigen}), the left eigen vectors
and right eigen vectors are not really independent.  It is possible to choose
the left eigenvectors ${\tilde x} = (\tilde \mu, \tilde \epsilon) =
(\epsilon^\dagger, - \mu^\dagger)$.  The orthonormal condition then holds between
the left and right eigen vectors.  In particular, the eigen modes can be
normalized according to
\begin{equation}
\epsilon^\dagger \epsilon + \frac{i}{\omega} \epsilon^\dagger\! A \epsilon = 1.
\end{equation}
Since the matrix on the right-hand side of Eq.~(\ref{eq-eigen}) is not
anti-hermitian, there is no guarantee that the frequencies $\omega$ will be
real, but the eigenvalues always come in pairs, $\pm \omega$.  We take only
$\omega>0$ modes.  With these choices of the eigen modes, displacement and
momentum operators can be taken in the standard second quantization form,
\begin{eqnarray}
u_l &=& \sum_{k} \epsilon_k e^{i {\bf R}_l \cdot {\bf k}} \sqrt{\frac{\hbar}{2\omega_k N}}\; a_k + {\rm h.c.},\\
p_l &=& \sum_{k} \mu_k e^{i {\bf R}_l \cdot {\bf k} } \sqrt{\frac{\hbar}{2\omega_k N}}\; a_k + {\rm h.c.},
\end{eqnarray}
where $k=({\bf k},\sigma)$ specifies the wavevector as well as phonon
branch, $a_k$ is the annihilation operator, and h.c.\ stands for hermitian
conjugate.  The momentum and displacement polarization vectors are related by,
e.g., $\mu = -i\omega \epsilon + A\epsilon$.  We can verify that the canonical
commutation relations are satisfied, $[u_l, p_{l'}^T] = i\hbar \delta_{l,l'}
I$, and $H= \sum_{k} \hbar \omega_k (a_k^\dagger a_k + 1/2)$.

Based on a definition of the local energy density and the continuity equation
for energy conservation, an energy current density can be defined as
\cite{hardy63,sheng06,kagan08},
\begin{equation}
J^c = \frac{1}{2V} \sum_{l,l'} (R_l^c\! -\! R_{l'}^c) u^T_{l} K_{l,l'} \dot{u}_{l'},
\end{equation} 
where the index $c=x$, $y$, or $z$ labels the cartesian axis, $V$ is the total
volume of $N$ unit cells.  The components of the current density vector can be
expressed in terms of the creation/annihilation operators. Ignoring the $a\,a$
and $a^\dagger a^\dagger$ terms which vary rapidly with time, one obtains
\cite{kagan08}
\begin{equation}
J^c = \frac{\hbar}{4V} 
\sum_{k,k'} \left( \sqrt{\frac{\omega_k}{\omega_{k'}}} + \!
\sqrt{\frac{\omega_{k'}}{\omega_k}} \right) 
\epsilon_k^\dagger \frac{\partial D({\bf k})}{\partial k^c} \epsilon_{k'}\,
a_k^\dagger a_{k'} \delta_{{\bf k},{\bf k}'}.
\end{equation}

The thermal conductivity tensor can be obtained from the Green-Kubo formula
\cite{mahan00},
\begin{equation}
\kappa_{ab} = \frac{V}{T} \int_0^{\beta\hbar}\!\!\!\! d\lambda 
\int_0^\infty\! dt\, \bigl\langle J^a(-i\lambda) J^b(t) \bigr\rangle_{\rm eq},
\label{eq-GK}
\end{equation}
where $\beta=1/(k_BT)$, the average is over the equilibrium ensemble with Hamiltonian $H$.  The
time dependence of the annihilation operator is trivially given by $a_k(t) =
a_k e^{-i\omega_k t}$.  This is also true if $t$ is imaginary.  Substituting
the expression $J^c$ into Eq.~(\ref{eq-GK}), using the result
\begin{equation}
\langle a_i^\dagger a_j a_k^\dagger a_l \rangle_{\rm eq} = 
f_i f_k \delta_{ij}\delta_{kl} + f_i (f_j+1)\delta_{il} \delta_{jk},
\label{eq-aaaa}
\end{equation}
where $f_i = (e^{\beta\hbar \omega_i} - 1)^{-1}$ is the Bose distribution
function, we obtain
\begin{eqnarray}
\kappa_{ab} &=& \frac{\hbar}{16VT} \sum_{{\bf k}, \sigma, \sigma'}
\frac{e^{\hbar(\omega'-\omega)\beta}-1}{\omega'-\omega} \
\frac{1}{\eta - i (\omega- \omega')} \times \qquad\nonumber \\
       && F_{\sigma'\sigma}^a({\bf k})
        F_{\sigma\sigma'}^b({\bf k}) f(\omega') \bigl( f(\omega) + 1\bigr),
\label{eq-main1}
\end{eqnarray}
where the $F$ function is defined as
\begin{equation}
F_{\sigma\sigma'}^a({\bf k}) =
\left( \sqrt{\frac{\omega}{\omega'}} + 
\sqrt{\frac{\omega'}{\omega}} \right) 
\epsilon^\dagger \frac{\partial D({\bf k})}{\partial k^a} \epsilon'.
\label{eq-main2}
\end{equation}
To simplify notations, we have suppressed indices, e.g.,
$\omega=\omega_\sigma({\bf k})$, $\epsilon' = \epsilon_{\sigma'}({\bf k})$.
We have added a damping term $e^{-\eta t}$ when integrating the oscillatory
factor.  The diagonal element of $F$ is related to the group velocity,
$F^a_{\sigma\sigma}({\bf k}) = 2\, \partial \omega^2_k/\partial k^a$.  The
off-diagonal elements are in general not zero.  The first term in
Eq.~(\ref{eq-aaaa}) factors into two independent summations which does not
contribute to $\kappa_{ab}$ due to symmetry of $\omega_{\sigma}({\bf k})$ with
respect to the wavevector ${\bf k}$.  Equation~(\ref{eq-main1}), together with
the definition (\ref{eq-main2}), is the main result of this paper.

We make some general comments on Eq.~(\ref{eq-main1}).  The first and last
factors inside the summation sign can be combined, $(e^{\beta \hbar
(\omega'-\omega)}-1) f' (f+1) = f -f'$. Written in this way, the equation
resembles the Landauer formula for ballistic transport.  The second factor
makes the conductivity diverge in the form $1/\eta$ unless the leading term in
an expansion in the damping factor $\eta$ happens to be zero.  The diagonal
elements $\kappa_{aa}$ indeed diverge to infinite. This is expected, as the
system is ballistic consisting of independent oscillating modes. There is no
intrinsic scattering mechanism in the system.

The off-diagonal elements do not diverge if the system is isotropic in the
sense that $\kappa_{ab}$ is independent of the choice of the coordinate axis.
In this special case, the off-diagonal elements are antisymmetric and odd in the
magnetic field $h$, $\kappa_{ab}(h) = -\kappa_{ba}(h) = \kappa_{ba}(-h)$, consistent
with the Onsager relation.  This property does not hold for arbitrary
anisotropic systems.  We argue that in the isotropic case,
Eq.~(\ref{eq-main1}) is physical and is the correct prediction for the Hall
thermal conductivity.

Even in the isotropic case, the off-diagonal term is zero unless reflection
symmetry is broken.  More precisely, if there exists an orthogonal
transformation independent of ${\bf k}$ such that $S D S^T = D$, $SAS^T = -A$,
then $\kappa_{ab} = 0$ for $a \neq b$.  The physical meaning of this symmetry is
clear.  If we look the system in a mirror, since $D$ is the same and $A$ flips
a sign, but the physics must be invariant, we should have $\kappa_{ab}(D,A) =
\kappa_{ab}(D,-A)$. But $\kappa_{ab}(D,A)$ must be an odd function in $A$.  So we must
have $\kappa_{ab} = 0$, $a \neq b$.  This property should be quite general,
independent of the models used. As an example of systems with vanishing
off-diagonal thermal conductivity, we can cite a square lattice (or cubic
lattice) with only the nearest neighbor coupling with a dynamic matrix which is
diagonal.

\begin{figure}
\includegraphics[width=\columnwidth]{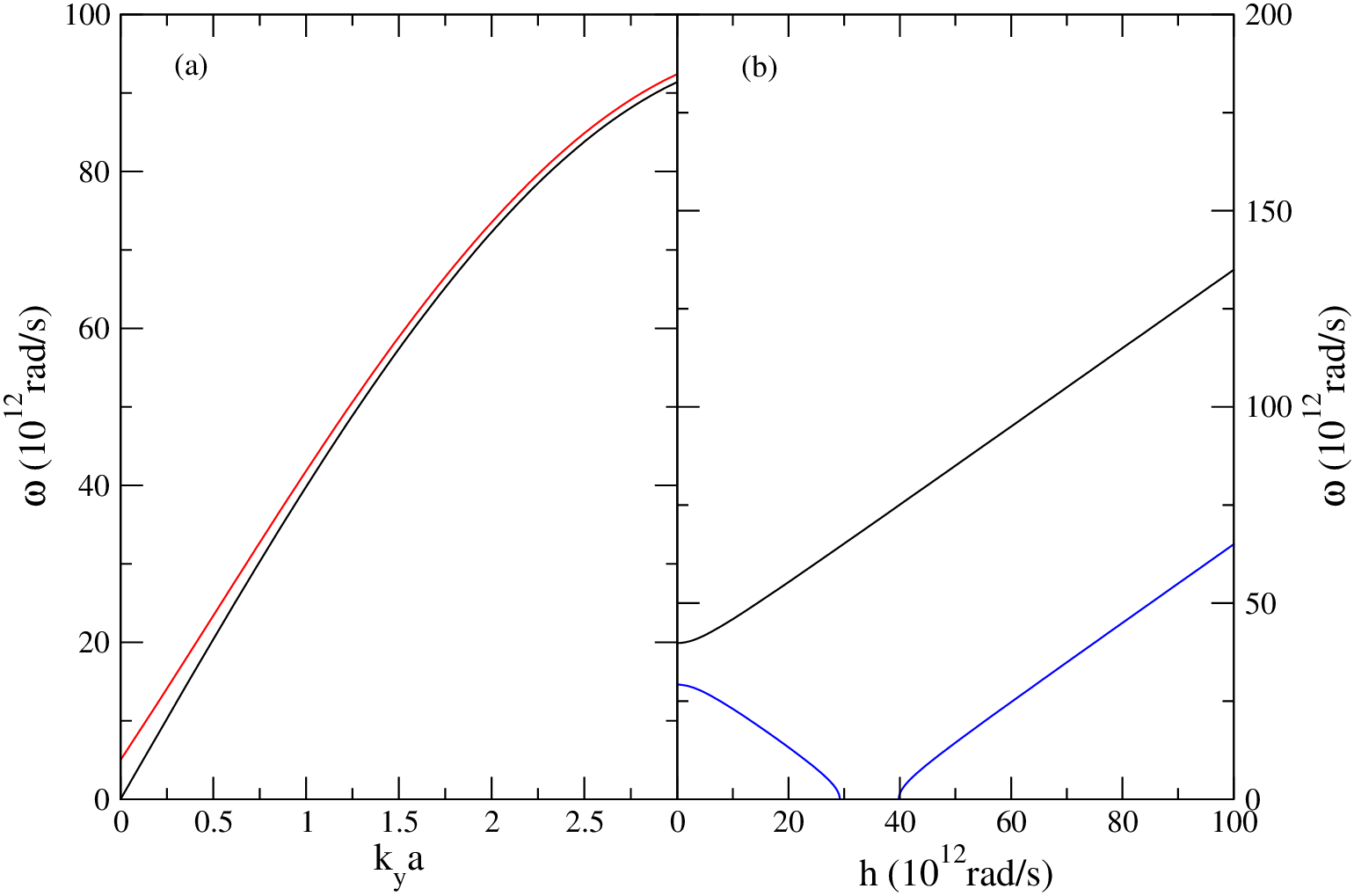}%
\caption{\label{fig1}Phonon dispersion relation of a triangular lattice.  (a)
The angular frequency of longitudinal mode as a function of $k_ya$ with
$k_x=0$.  The bottom curve is $h=0$; top curve is $h=5\times
10^{12}\,$rad$\,$s$^{-1}$.  (b) The frequency as a function of $h$ at a fixed
wavevector ${\bf k}a = (0,1)$. The top curve is the longitudinal mode, the
bottom (broken) curve is the transverse mode.}
\end{figure}

In the following, we present numerical results based on Eq.~(\ref{eq-main1}).
But first, we discuss some interesting features of the phonon dispersion when
the Raman interaction term is turned on.  In Fig.~\ref{fig1}, we show results
for a triangular lattice with only the nearest neighbor couplings.  The
coupling matrix between two sites is such that the longitudinal spring
constant is $K_L = 0.144\,$eV/(u\AA$^2$) and the transverse spring constant
$K_T$ is 4 times smaller.  The unit cell lattice vectors are $(a,0)$ and
$(a/2, a\sqrt{3}/2)$ with $a=1\,$\AA.  This choice gives longitudinal and
transverse sound velocity $3981\,$m/s and $1921\,$m/s, respectively,
comparable to typical experimental values.  At the $\Gamma$-point, the effect
of the interaction is to shift the frequencies from $\omega_0$ to $\omega_0
\pm h$, for both the acoustic modes and optical modes (if any).  In
particular, the acoustic modes develop a gap from 0.  Away from the
$\Gamma$-point, the corrections are of order $h^2$.  Due to the interaction,
some modes have imaginary frequencies and are no longer stable.  This is very
pronounced for the transverse modes for large $h$, see Fig.~1(b).  The system
can be stablized, at least for small $h$, by adding a small onsite potential
(which, of course, breaks the translational invariance of the lattice).  If we
change the model to use velocity $\dot{u}$ instead of the conjugate momentum
$p$ in the interaction term, an onsite term of magnitude $h^2$ is generated
naturally. In such a model, these instabilities do not appear.  However, there
is no good reason to use $u^T\! A \dot{u}$ instead of $u^T\! A p$ other than
the above observation.

\begin{figure}
\includegraphics[width=0.95\columnwidth]{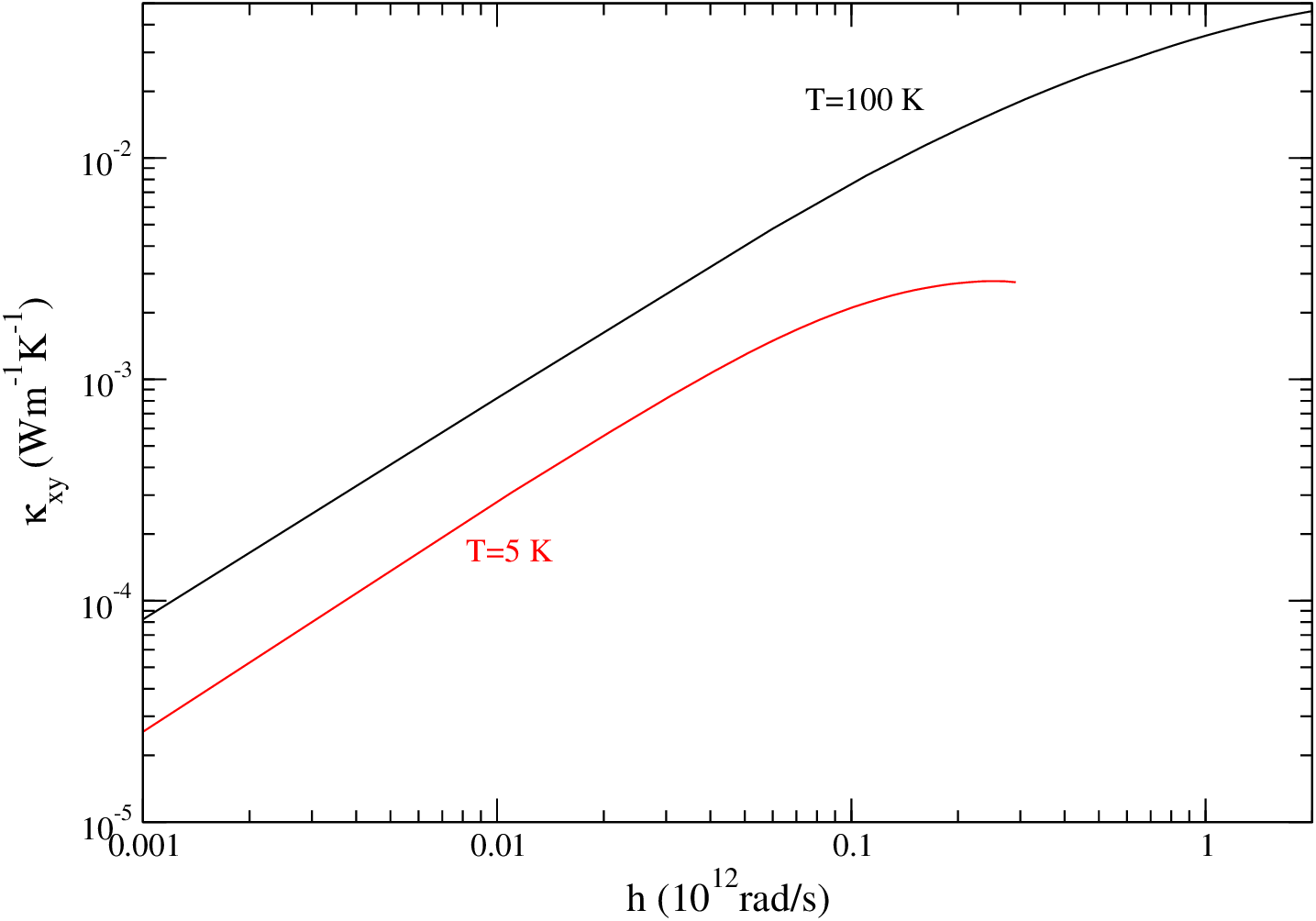}%
\caption{\label{fig2}Thermal Hall conductivity as a function of the coupling
$h$ for fixed temperatures $T=5\,$K and $100\,$K, respectively.}
\end{figure}

In Fig.~2, we give the off-diagonal thermal conductivity $\kappa_{xy}$ of the
triangular lattice (assuming $1\,$\AA\ thick) as a function of $h$ for two
different temperatures, $T=5\,$K and $100\,$K. It is clear that, for small
$h$, the dependence of $\kappa_{xy}$ on $h$ is linear.  For large $h$, the growth
becomes weaker than linear.  For very large $h$ (not shown), due to the
instability, $\kappa_{xy}$ becomes rather singular, and can even become negative.
This range of parameters is not physical.  In computing the results of Fig.~2,
we have added a small onsite value of order $10^{-6}K_L$. The results are
sensitive for this onsite value only for large $h$, but are nearly independent
of the onsite value for small $h$.

\begin{figure}
\includegraphics[width=1.00\columnwidth]{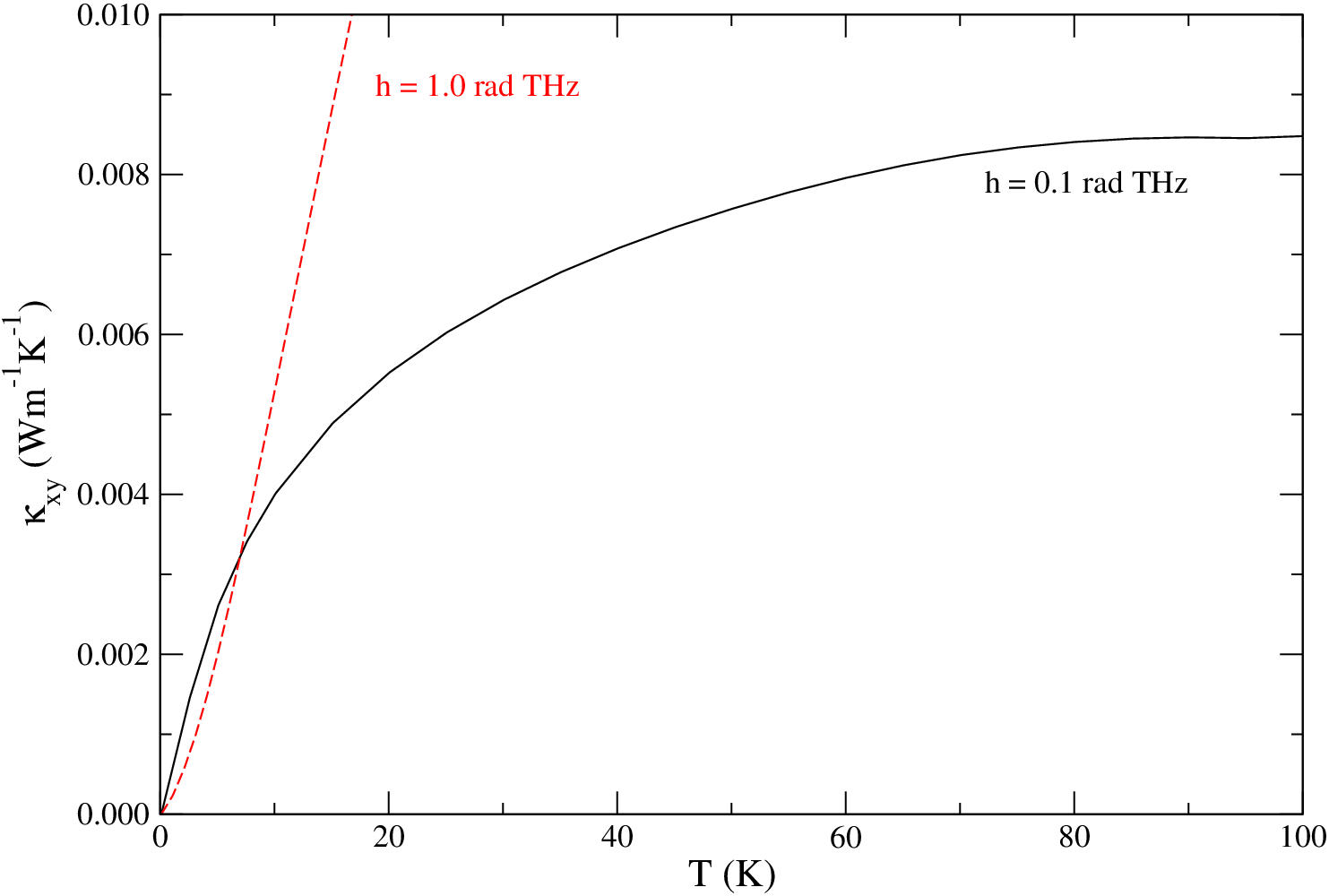}%
\caption{\label{fig3}Thermal Hall conductivity $\kappa_{xy}$ as a function of the
temperature $T$ for fixed coupling $h=10^{11}\,$rad/s (solid line) and 
$10^{12}\,$rad/s (dotted line).}
\end{figure}

In Fig.~3, we display the temperature dependence of the off-diagonal thermal
conductivity $\kappa_{xy}$.  It is seen that $\kappa_{xy}$ saturates at about 100 K at
$h=0.1\,$rad$\,$THz.  At low temperatures, $\kappa_{xy}$ decreases with temperature
approximately linearly.  However, due to a complicated effect of $h$ to the
dispersion relation, it appears that $\kappa_{xy}$ has a faster fall off than
linear.

We comment on experimental data \cite{strohm05,inyushkin07} in comparison with
our numerical results. We have not taken into account the specific lattice
structure and atomic details used in experiments.  A quantitative comparison
is not possible.  However, the phonon model parameters are comparable to real
systems by matching the sound velocities.  The most uncertainty in a comparison is
the coupling $h$.  The experimental value for $\kappa_{xy}$ at $T=5.13\,$K and
magnetic field $H=3\,$T is $2.0\times 10^{-5}$Wm$^{-1}$K$^{-1}$
\cite{inyushkin07}.  This is consistent with a very small value of
$h=10^{-3}\,$rad$\,$THz.  Although the diagonal element $\kappa_{xx}$ $(=\kappa_{yy})$
diverges to infinite in our theory, we can choose a finite $\eta$ in
Eq.~(\ref{eq-main1}) to mimic a finite phonon life time.  We find a very weak
dependence of $\kappa_{xx}$ on $h$.  On the scale of $h \sim 1\,$rad$\,$THz,
$\kappa_{xx}$ is nearly a constant.  By matching the experimental value of order
0.5 Wm$^{-1}$K$^{-1}$, we can infer a mean free path $\ell = c/\eta \approx
10^3\,$\AA\ (where $c$ is sound velocity), which appears a bit too small given
the very low temperatures in experiments.

It is interesting to compare the present treatment with that of nonequilibrium
Green's function (NEGF) approach in Ref.~\cite{lifa09}.  The qualitative
features are in agreement, such as the vanishing phonon Hall effect on square
lattice.  In NEGF approach, the leads are modeled explicitly.  It was assumed
that leads do not have the spin-phonon interaction.  This has the advantage of
stablizing the system, even though the spin-phonon system represented by the
Hamiltonian $H$, Eq.~(\ref{eq-model}), may be unstable as a bulk system.  NEGF
deals with very small systems in practice.  Some of the oscillatory behavior,
perhaps of a reflection of the wave nature, is not found here.  The present
theory is more suitable for comparison with experiments which were done on
samples of mm scale.

Another point is the role of nonlinear interactions. The phonon-phonon and
spin-phonon interactions will produce a finite life time for the phonons,
rendering a finite thermal conductivity tensor for all components.  We expect
that, if there is a systematic expansion in terms of the phonon life-time or
in terms of the interaction strength, our main result, Eq.~(\ref{eq-main1}),
should be the leading contribution.  The interaction should give only small
corrections.

In summary, we have presented a theory of phonon Hall effect based on a
ballistic lattice dynamic model.  It is shown that the phonon Hall effect can
be present, provided that the system does not possess a reflection symmetry.
This is different from Ref.~\cite{sheng06}, which does not suggest this
subtlety.  Since the Hamiltonian is quadratic in the dynamic variables, a
perturbative treatment is not necessary.  In fact, it fails near the
$\Gamma$-point.  We have given numerical results on a simple two-dimensional triangular
lattice and the qualitative features are the same for all lattices in two and
three dimensions.  In particular, it is not necessary that the system has
degenerate phonon modes when $h=0$.  When more elaborate model is known (e.g.,
from a first-principles calculation), the current theory can be applied to
more realistic systems.

This work is supported in part by a research grant of National University of Singapore
R-144-000-173-101/112.

\end{document}